\documentclass[10pt,letterpaper]{article}
\usepackage{epsfig}

\begin{document}

\title{\large \bf DIP: Disruption-Tolerance for IP}
\author{Michael Neufeld \\
BBN Technologies \\
10 Moulton St. \\
Cambridge, MA 02138 \\
mneufeld@bbn.com}

\maketitle

\begin{abstract}
{\em Disruption Tolerant Networks} (DTN) have been a popular subject
of recent research and development. These networks are characterized
by frequent, lengthy outages and a lack of contemporaneous end-to-end paths.
In this work we discuss techniques for extending IP to operate more
effectively in DTN scenarios. Our scheme, Disruption Tolerant IP (DIP),
uses existing IP packet headers, uses the existing socket API for applications,
is compatible with IPsec, and uses familiar Policy-Based Routing techniques
for network management.
\end{abstract}

\section{Introduction}
\label{sec:intro}
Of late, {\em Disruption Tolerant Networks} (DTN) 
have been a subject of research and development\cite{dtn:rfc,5050:rfc}.
These networks are characterized by frequent, potentially lengthy outages
as well as a lack of contemporaneous end-to-end paths.
\begin{figure}
\centerline{
\epsfig{file=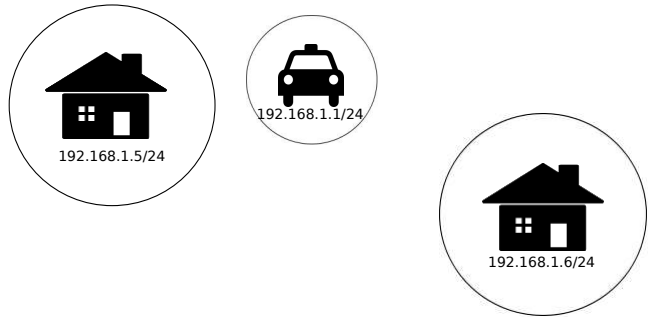,width=0.7\columnwidth}
}
\caption{\footnotesize This figure shows a car acting as a ``data mule,''
ferrying information between the two houses. There is never a complete
path between the two houses at any given time, only a path between
one of the houses and the car at any given time.}
\label{figure:datamule}
\end{figure}
Figure~\ref{figure:datamule} illustrates a typical ``data mule''
network scenario where DTN may be applied. Standard IP routing and forwarding
fails to transfer packets between the two houses because there is never a
complete end-to-end path between them. A DTN can exploit the periodic
connection provided by the car to transfer data between the two houses.
Interplanetary networks have similar connectivity patterns:
orbital mechanics result in ``scheduled'' links between
nodes and non-connected network topologies.
These disrupted networks present challenges that standard IP networking
is not well-equipped to handle. 
TCP backoff in the face of delayed and dropped packets and its reliance
on bidirectional end-to-end communication make it unsuited for
disrupted networks, though {\em Performance Enhancing Proxies}
\cite{3135:rfc} or other schemes\cite{dttcp} can bolster
TCP in disrupted networks. UDP delivery
semantics {\em are} compatible with disrupted networks but
IP forwarding drops packets immediately when no next hop is available.
Any forward progress made by packets over sporadic links is abandoned.
Existing applications, {\em e.g.} {\em Situational Awareness} tools used
to coordinate emergency rescue and military tactical operations, could benefit
from allowing packets to remain in queues until a route comes up instead
of dropping packets that do not have immediate routes.

Operation in disrupted networks has primarily been an application layer issue,
though the {\em Bundle Protocol}\cite{5050:rfc} provides disruption
tolerance as a basic network service. However, using the Bundle Protocol
requires updating legacy applications or incurring the overhead of 
tunneling within bundles. In this work we will show how IP may be
modified to better handle disrupted networks without introducing an entirely
new protocol, and how these modifications to IP fit in with the Bundle Protocol.

We will begin by broadly splitting disrupted networks into two categories:
networks that experience disruption {\em without} changes to endpoint
identifiers (static disruption) and those that experience disruption
{\em with} changes to endpoint identifiers (dynamic disruption). Static
disruption is generally easier to handle than dynamic disruption because
static disruption does not require tracking endpoint identifier changes
during the course of a communication session. Unfortunately the
term {\em identifier} has been used to mean different things within
the context of networks. In this work we will use {\em identifier} to
mean a combination of {\em name} and {\em location} as
defined by the {\em Uniform Resource Identifier}\cite{3986:rfc} (URI).
Other relevant work, {\em e.g.} the {\em Host Identity Protocol}
\cite{4423:rfc} (HIP) uses {\em identifier} in the same way that
we use {\em name}: a location-independent tag that corresponds to an entity.

An ``endpoint identifier'' is an abstract entity, referring to a producer
or consumer of information. An actual network system must have specific
entities that constitute endpoints and ways of identifying those entities.
IP networks use  {\em applications} (identified by 16-bit port numbers)
connected to {\em network interfaces} (identified by IP addresses) as endpoint
identifiers. IP addresses inherently combine name and network location,
meaning that mobility often causes a change in endpoint identifier.

Despite this tight binding of endpoint identification to network
location, there are a variety of static disruption scenarios
for IP networks. A straightforward class of statically disrupted
IP networks have fixed topology and intermittent
links. Intermittency may occur because of external
interference or deliberate scheduling policy. External interference
is usually driven by environmental factors, {\em e.g.} RF interference
and atmospheric conditions for freespace RF and optical networks or
passing boat noise for underwater acoustic networks.
Policy-based intermittency is frequently driven by {\em cost},
{\em e.g.} operating a solar-powered network primarily
during sunny daylight hours or exploiting inexpensive phone rate periods
in UUCP networks.

Less intuitively, {\em ad hoc} and mobile wireless IP networks may also
qualify as ``static'' disrupted networks.
{\em Ad hoc} routing protocols often operate using 
flat, static IP address assignments, and {\em Mobile IP}\cite{mobileip:rfc}
provides the illusion of an unchanging IP address despite changes
in actual network connection point. The data mule example in
Figure~\ref{figure:datamule} exhibits static disruption because of its
fixed IP address assignment. Dynamic disruption in IP networks occurs
whenever node mobility is accompanied by a change in IP address, {\em e.g.}
a mobile wireless node configured using DHCP and moving between
access points connected to different subnets.

In this work we will present {\em Disruption Tolerant IP} (DIP): a system
augmenting IP to handle static disruption. Our scheme maintains the
existing IP packet format, is directly compatible with the existing
socket interface, uses familiar Policy-Based Routing techniques for 
network management, and is compatible with IPsec. 

\section{DIP: Disruption-Tolerant IP}
DIP operates along the same lines as conventional
congestion queueing, but is more complex. Like congenstion queueing,
DIP operates directly in conjuction with IP forwarding.
We will examine the specific details of how DIP integrates into
existing IP forwarding after we outline the tasks that DIP
must perform.

In order to achieve static disruption
tolerance DIP must perform the following five tasks:
acquire disrupted packets, track packet longevity and drop
expired packets, track disruption periods and available routes,
deliver disrupted packets when routes become available,
and manage limited storage space. These tasks will be described in
greater detail in the following sections.

\subsection{Packet Longevity}
\begin{figure}
\centerline{
\epsfig{file=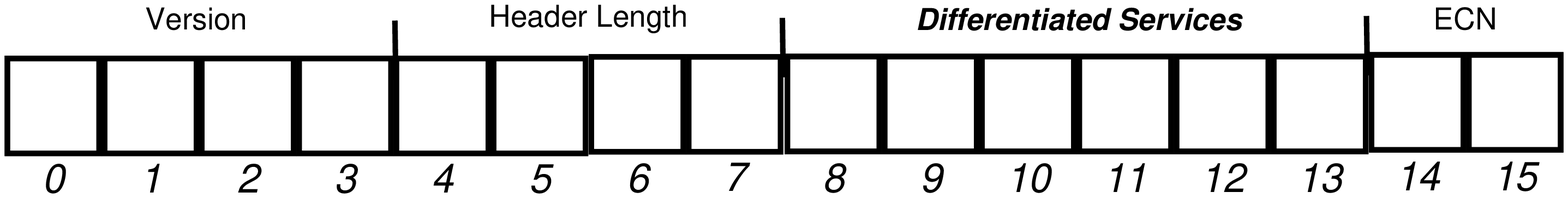,width=0.9\columnwidth}
}
\caption{\footnotesize The DSCP field is bits 8 through 13 in the IP header,
following the {\em IP Version} and {\em Header Length} fields, and followed
by the Explicit Congestion Notification ({\em ECN}) bits.}
\label{figure:dscp_field}
\end{figure}

In DIP, packet longevity is defined to be the length of
time that a packet may be allowed to spend ``at rest'' in queues.
This simple definition makes minimal demands on clock synchronization and
allows small numbers to be used to represent lifetimes. 
DIP utilizes the {\em Differentiated Services Code Point} (DSCP) IP header
field (often referred to as the IP {\em Type of Service}, or ToS, field)
to express packet lifetime. The DSCP field is near the beginning of the 
IP header, as illustrated in Figure~\ref{figure:dscp_field}.

The DSCP field is suitable for expressing
packet longevity for three primary reasons: (1) there are DSCP bits available
for use and (2) the DSCP bits may be safely changed by intermediate
routers during transit, and (3) the DSCP bits may be set by the
originating application through the standard socket API.
RFC 2474\cite{2474:rfc} describes the mechanics
of using DSCP bits for both IPv4 and IPv6, and in section 6 defines two pools
of 16 DSCP values for local/experimental use. These two pools provide
up to five bits within each IP header that may be used to describe
packet longevity and whatever additional packet priority is required for
traffic shaping. The {\em REDTIP} \cite{redtip} system also proposes
using DSCP bits for delay tolerance, but does not delve as deeply
into the details as we will for DIP.

\begin{figure}
\centerline{
\epsfig{file=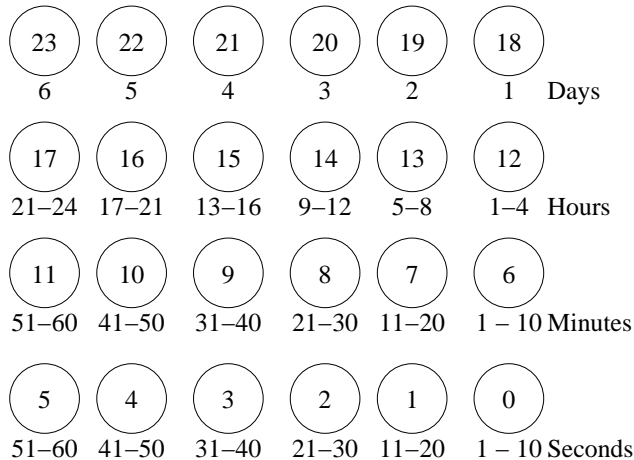,width=0.7\columnwidth}
}
\caption{\footnotesize A ``timer wheel'' structure for coarsely tracking
packet lifetime. Packets are initially placed in bins corresponding to
longevity category. A periodic timer tracks the shortest required granularity
(10 seconds in this example). Each timer tick flushes bin 0
and rotates bin labels to reflect new packet lifetimes. As packets
are dequeued their IP DSCP bits are marked with their current
lifetime.}
\label{figure:packet_timer_wheel}
\end{figure}

Using the DSCP bits to specify exact longevity seems
wasteful, {\em e.g.} using seconds as a base unit only allows
a packet longevity of 32 seconds and uses all available DSCP
values. By coarsely tracking longevity a useful disruption tolerance may be
achieved while maintaining a compact representation.
Four values (two bits) could specify
longevities on the order of seconds, minutes, hours, and days.
A scheme along the lines of a {\em hierarchical timer wheel}
could then be used to track and update packet longevity 
efficiently. Figure~\ref{figure:packet_timer_wheel} illustrates
this scheme.
An arriving packet is placed in the bin tracked by the top
end of its longevity category. In Figure~\ref{figure:packet_timer_wheel}
a packet in the ``hours'' category would be placed in bin marked ``17'' upon
arrival. A periodic timer tracks the shortest required granularity for
a given system. Each time
the periodic timer fires, the packets in bin 0 are flushed and the bin
labels rotated as needed. In Figure~\ref{figure:packet_timer_wheel}
each 10 second time interval would shift all of the bins in the
``seconds'' category, {\em i.e.} bin 0 would be flushed and remarked as bin 5,
bin 5 remarked as bin 4, bin 4 as bin 3, bin 3 as bin 2, {\em etc.}
On every 60th rotation ({\em i.e.} at 10 minutes) the rotation would be
extended to include bin 11, on every 1440th rotation ({\em i.e.} at 4 hours)
extended further to include bin 17, and finally every 8640th rotation
({\em i.e.} one day) to include bin 23. As packets exit a particular bin,
their IP ToS is marked to reflect their current lifetime category.

\begin{figure}
\centerline{
\epsfig{file=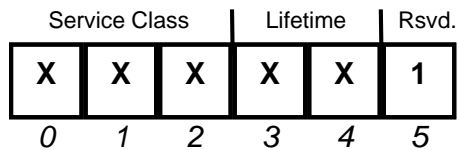,width=0.5\columnwidth}
}
\caption{\footnotesize DSCP bitfield usage for DIP. Bits 0-2 provide
8 possible service classes and bits 3 and 4 provide four packet lifetime
categories. Bit 5 must be set to 1 in order
to match the DSCP pools designated for experimental use by RFC 2474.}
\label{figure:dscp_bits}
\end{figure}

DIP does not {\em guarantee} timely expiration of packets in all cases.
A packet could jump from node to node at a fast enough rate to
avoid ever having its lifetime decremented below its initial level.
In such cases the IP TTL field will serve to limit the packet's time
in the network. Furthermore, DIP only
provides ``best effort'' packet lifetime guarantees: at any point a
DIP system may opt to drop packets due to a lack of storage space.
As with normal IP QoS unexpected results
may arise if use of the DSCP bits is not consistent across routers.
As a starting point for experimentation, we propose using two bits
for longevity and three bits for service class
as shown in Figure~\ref{figure:dscp_bits}.
The service class values used in DIP may align with the standard
precedence values for IP if compatibility with existing per-hop router
behavior is desired. Alternatively
a break may be made with existing IP precedence and the
service class bits used to specify alternate queueing policies.
For example, relative importance could be used instead of precedence.
Unimportant packets could have long lifetimes but placed in short queues
to limit their resource consumption.

\subsection{Managing Disruption and Disrupted Packets}
Effectively managing disruption and disrupted packets requires more
than a simple queue for disrupted packets. Routing must notify
DIP when new routes become available to destinations, and DIP must
be able to quickly locate packets for those destinations. Furthermore,
DIP must have some reasonable packet reinjection policy to avoid
overloading network links when reinjecting disrupted packets while still
taking good advantage of links as they appear. DIP must also manage the
storage space required for disrupted packets. Packets must be selectively
discarded when space limitations arise, and efficient duplicate detection
used to prevent packet retransmissions from clogging packet queues.

\subsubsection{Topology Updates and Packet Indexing}
Routing updates
come from two sources: {\em neighbor discovery} for directly reachable
hosts and {\em routing} for multihop hosts. DIP may utilize
information from either or both of these sources. Neighbor
discovery augments coarse (but scalable) subnet-level knowledge
with more precise host-level knowledge for local networks. In IPv4
neighbor discovery is implemented via ARP\cite{arp:rfc}, and IPv6
has specialized neighbor discovery\cite{4861:rfc}.

For hierarchical IP routes DIP utilizes a trie to index packet queues
holding disrupted packets by subnet. Routes arriving from neighbor
discovery will generally be in the form of host routes on a single subnet.
A simple implementation of DIP could maintain packet queues indexed by
individual destinations, though such a scheme would quickly be overwhelmed
if there are a large number of active neighbors. Hashing by destination
(similar to the per flow hashing used by Stochastic Fair Queueing),
provides the ability to trade off the number of queues required against
the possibility of ``head of line'' blocking.

Routing protocols may range from traditional IP static routing to explicitly
disruption/mobility aware protocols. DIP in conjunction with
an {\em ad hoc} routing protocol allows exploitation of
the enhanced neighbor discovery schemes utilized by {\em ad hoc}
routing protocols and provides a technique for packet queue management.
Explicitly disruption-tolerant
routing protocols, {\em e.g.} {\em Zebranet}\cite{zebranet}, could
feed routing information into DIP, as could ``scheduled'' routing
protocols, {\em e.g.} space networks whose connectivity depends on
well-calculated orbital mechanics.
DIP forwarding may even integrate input from {\em multiple} routers
to determine where to next send packets. The details of interfacing with
multiple routers and how best to assimilate and act upon information from
them is currently an open research topic.

\subsubsection{Reinjecting Packets}
Reinjecting packets into the network once routes become available
requires careful design. Immediately sending all available packets
could easily overwhelm the network, so some combination of
backpressure and shaping is required for effective network utilization.
Packets may be available for different destinations,
at different priority levels, and with different lifetimes. Fortunately
routing and forwarding techniques for fair queueing, traffic
shaping, and differentiated packet priorities already exist, providing
mechanisms for such tasks. Existing {\em Policy-Based Routing} (PBR)
schemes may be extended to include DIP. Policy-Based Routing allows
packet queueing disciplines and dropping policies to be set based
on various properties of packets, {\em e.g.} source address, destination
addresses, incoming network interface, and the DSCP field. By treating
the disruption queue as a ``network interface'' DIP can present
a management interface that allows specific queueing policies to 
be set for packets entering and leaving the disruption queue.
Advanced routing and flow control protocols may also have
information about expected link capacity/length of contact time that
could be used to appropriately shape outgoing traffic.

\subsubsection{Dropping Duplicate Packets}
Allowing packets to remain in the network for a long time emphasizes
packet retention and dropping policies. Duplicate packets may
wastefully clog queues and hamper delivery of non-redundant packets.
Efficient duplicate detection hinges on {\em digesting} and
{\em indexing} packets efficiently. The {\em IP identifier}
field in combination with the source and destination of each
packet is a simple choice for a packet digest. Unfortunately
the limited size of the IP identifier combined with potentially
long time scales and high capacity links means that we must
contend with collisions. A key observation to make is that
the packet tracking requirements of DIP are similar to those
of {\em packet traceback} systems that track many individual
packets over extended periods of time. Given this similarity
DIP can leverage techniques from the
{\em Source Path Isolation Engine}(SPIE)\cite{spie} system to perform
duplicate detection.

SPIE uses a packet digest that includes
immuatble IP header fields and the first eight bytes of the payload
and {\em bloom filters}\cite{netbloom} to index the packet digests.
DIP requires slight variations on the techniques used in SPIE. For indexing SPIE
uses regular bloom filters that do {\em not} support deletion.
DIP must track duplicates of packets stored within its queues,
not necessarily tracking every packet ever sent through its queues.
Baseline bloom filters do not allow element deletion, so DIP must
use {\em counting} bloom filters\cite{netbloom} that do permit deletions.
The bloom filters used in DIP must be tuned to provide an acceptable
false positive rate when DIP queues are full.

\subsubsection{Full Packet Queues}
DIP must behave sensibly when packet queues fill and packets
must be dropped. Simple tail dropping is an obvious strategy
for DIP to support. However, an application that values newer
packets over old may benefit from queues that drop from the
{\em head}, preferentially keeping more recently received packets over
old. For example, a situational awareness system may find old
updates of interest but prefer to obtain more recent information first.
Schemes that continually winnow packets before queues are filled
({\em e.g.} RED\cite{red}) may have some utility in disrupted network,
though further study is required on this subject. With these
factors in mind, DIP will support packet queues that drop from the
tail, the head, as well as active queue management schemes that
preemptively drop packets before queues are full.

\section{Integrating With Forwarding and Routing}
\begin{figure}
\centerline{
\epsfig{file=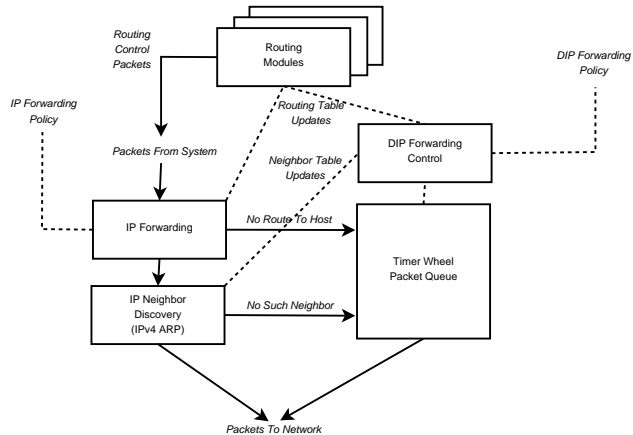,width=0.7\columnwidth}
}
\caption{\footnotesize Integrating DIP into IP requires hooks into routing,
forwarding, and neighbor discovery (ARP in IPv4 systems). The routing
protocol(s) in use may be traditional IP routing, ad hoc routing,
or explicitly disruption aware routing protocols.}

\label{figure:staticdipdetail}
\end{figure}

Figure~\ref{figure:staticdipdetail} shows how DIP hooks into existing IP 
systems.
The primary low-level points at which DIP must hook into the existing IP
stack are within IP forwarding and Neighbor Discovery/ARP. For nodes
operating as simple clients with a default route, the most important
hook is the one into ND/ARP. When a default route is set, {\em all}
packets by definition have a route: the only way to detect disruption
is within the context of ND/ARP on the local link.

Acquiring packets from ARP and/or ND may require modification
of the underlying network stack. Furthermore, IP ARP and ND may
not detect neighbor changes rapidly enough to prevent significant packet
loss. At the very least some tuning of ARP and ND will be required,
and potentially the employment of a new neighbor discovery scheme
better suited to detecting link outages, {\em e.g.} using information
from the underlying MAC layer. For an initial prototype of our
system, we will use the {\em Click Modular Router}\cite{click:tocs00} 
to provide an easily modifiable and efficient IP forwarding stack.
For nodes that do {\em not} normally have a default route,
DIP may acquire packets by operating as a ``router'' on
a virtual network device and installing a default route to itself.

For systems using dynamic routing protocols, DIP must also acquire information
from IP routers. As a starting point, DIP can acquire routing information
via the forwarding table. If DIP requires more detailed access to the full
Routing Information Base ({\em RIB}) we will use the {\em XORP}\cite{xorp}
routing framework to provide a unified mechanism for interfacing with
routers. Configuring traditional IP routers to operate with DIP
will require some adjustment. For example, with DIP {\em static}
routing may be function well on scheduled/unreliable links.
As long as a reliable neighbor discovery protocol prevents sending
packets to unavailable neighbors, packets may be sent immediately
and travel until they hit a disruption point in the network. At
that point they may reside until the disruption is cleared or they expire.

DIP must also interact with IP fragmentation/reassembly. Typical IP
fragment reassembly timeouts are much shorter than the potential
packet lifetimes allowed by DIP. Furthermore the IPv4 ID field
used to collect fragments is only 16 bits and is likely to be reused
during DIP packet lifetimes. IPv6 extends the length of the IP ID field
to 32 bits but could still experience collisions when long packet
lifetimes are used, {\em e.g.} sending 1000 fragmented packets per second
would exhaust a 32 bit space in slightly under 50 days.
DIP can mitigate these problems by keeping packet fragments close together
in time. As long as there isn't a large delay between the delivery
of all the fragments for a particular packet {\em and} fragments
from different packets with the same IP identifier are not interspersed
IP fragmentation can function. It may even be worthwhile for DIP
to reassemble long lived IPv4 packets at each hop (IPv6 packets may
only be fragmented by the source), or at least make
sure that most or all fragments have been collected before forwarding.

In conjunction with this traditional IP prototype we will
also integrate DIP into an {\em ad hoc} routing protocol, {\em e.g.}
AODV\cite{aodv:rfc} or OLSR\cite{olsr:rfc} as well as an opportunistic
routing protocol, {\em e.g.} a routing scheme based on estimated
probability of future contact such as {\em Zebranet}\cite{zebranet}.

\subsection{DIP and the Bundle Protocol}
There is some overlap between DIP and the Bundle Protocol,
but the Bundle Protocol provides some key services that
DIP does not. First is the direct use of
{\em Application Data Units} (ADUs)\cite{alf} as a fundamental
network transport unit. DIP is limited to the maximum IP datagram
size for packets, {\em i.e.} 64KB for standard IP and 4GB if IPv6
jumbograms are allowed. Second, the Bundle Protocol attempts to
provide guarantees on data storage and delivery, allowing 
other network hosts to take custody of information from other
hosts, promising to keep that information in stable storage
and deliver it. DIP does not go beyond basic IP
packets and best effort delivery. Third, the Bundle Protocol
may use its endpoint identification scheme and hop-by-hop
routing to avoid dynamic disruption that IP cannot, {\em e.g.}
by using location-independent names for endpoints.
In exchange for its limitations, DIP has a much smaller and
simpler packet format, uses the existing socket API for
applications, is directly compatible with IPsec,
and may use familiar policy-based network management schemes.

DIP may be used to complement the Bundle Protocol as well as provide
disruption-tolerant services where the Bundle Protocol may not be available.
For example, resource constraints may make the bundle protocol too
expensive to operate at all nodes in a network, or bundles may have
to tunnel through an encrypted IP network that suffers disruption
but not have any bundle routing services. Bundles encapsulated in IP
packets with DSCP markings that approximate the longevity indicated in
the bundle headers could traverse a DIP fabric between Bundle Protocol
routing systems. Such bundles would receive a basic subset of desired
bundle services even when full bundle processing services aren't available.
For networks and applications that only require basic disruption tolerance
DIP may be used on its own, allowing lower overhead than the full
Bundle Protocol in such scenarios.

\section{Prior Work}
The Bundle Protocol\cite{5050:rfc} for DTN environments
\cite{dtn:rfc,dtn_arch_sigcomm} is a clear motivator for DIP, and the
success of the Bundle Protocol spurred the development of DIP.
The architecture for NIMROD mobility support\cite{2103:rfc}
also influenced our distinction between static and dynamic network disruption.
DIP is similar to the previously proposed REDTIP\cite{redtip}
system, though DIP provides more details about the specific use of
the IP DSCP bits as well as mechanisms for queueing disrupted packets.

\section{Conclusions and Future Work}
In this work we have presented techniques for making IP networks
more robust in DTN scenarios. These techniques may be used on their own,
integrated with the Bundle Protocol, or in conjunction with 
other network architectures ({\em e.g.} content-based networks)
that could benefit from disruption tolerance. We are currently
constructing a prototype implementation of DIP, and will
use this prototype to further explore routing and forwarding in
disrupted networks, particularly issues surrounding routing, flow control, and 
network management/control.

\bibliographystyle{unsrt}
\bibliography{sdip}

\end{document}